\documentstyle[prd,aps,eqsecnum,12pt]{revtex}

\title{
Lagrangian description of the fluid flow with vorticity in 
the relativistic cosmology}

\author{
Hideki Asada\footnote{Electronic address:
asada@phys.hirosaki-u.ac.jp} and Masumi Kasai\footnote{Electronic address:
kasai@phys.hirosaki-u.ac.jp} 
}
\address{
Faculty of Science and Technology, Hirosaki University, Hirosaki
036-8561, Japan
} 

\date{\today}

\begin{document}
\maketitle

\begin{abstract} We develop the Lagrangian perturbation theory in
the general relativistic cosmology, which enables us to take into 
account the vortical effect of the dust matter.  
Under the Lagrangian representation of the fluid flow, the propagation 
equation for the vorticity as well as the density is exactly solved. 
Based on this, the coupling between the density and vorticity is 
clarified in a non-perturbative way. 
The relativistic correspondence to the Lagrangian perturbation theory 
in the Newtonian cosmology is also emphasized. 
\end{abstract}

\section{Introduction}

The dynamics of fluids in the expanding universe is of great importance 
in the cosmology. 
For investigating such a dynamics, the Newtonian treatment is often used 
as a good approximation for the region $l/L \ll 1$, 
where $l$ is the scale of fluctuations of fluids and $L$ corresponds to 
the Hubble radius \cite{Peebles}.   
In the Newtonian cosmology, the Lagrangian perturbation theory 
in the dust universe has been developed for non-linear density fluctuations, 
up to the caustic formation \cite{Zeldovich,Buchert,BE,SK}. 
In order to take into account the relativistic correction to the Newtonian 
dynamics, the cosmological post-Newtonian approximation has been formulated 
\cite{Futamase,SA}. 
Furthermore, in the cosmological post-Newtonian approximation, 
the Lagrangian perturbation theory has been also discussed 
\cite{MT,TF}. 
Such a treatment beyond the Newtonian approximation may become important 
because of not only the theoretical interest but also the recent 
progress in the observational cosmology. 
For example, the Sloan Digital Sky Survey (SDSS) aims at the deep survey 
over several hundred Mpc \cite{SDSS}, in which it is not clear 
whether the Newtonian treatment is sufficient for such a large area. 

Relativistic theories of linear perturbations have been developed
\cite{Peebles,Bardeen,KS}.  
The second-order extension of relativistic perturbation theory has
been also developed \cite{Tomita}. 
However, they still depend on the assumption that the density
fluctuation is small. 
In order to overcome the drawback, several perturbative approaches 
for the nonlinear dynamics have been proposed, which are
 based on the fluid flow approach \cite{MPS}, 
 the gradient expansion method \cite{PSS,CPSS,SSC}, or 
 the relativistic Lagrangian approach \cite{Kasai,RMKB}. 
It is shown that the relativistic post-Zel'dovich approximation, 
a second-order extension of the relativistic Lagrangian approach, 
successfully express the non-linear evolution of the density contrast 
with higher accuracy than the conventional second-order theory 
\cite{MNK}. 

So far, however, these relativistic approaches are all restricted
within the irrotational dust.  
There has been no established way of
calculating non-linear evolution of the density fluctuation without
such limitation. Our 
paper is aimed to add to knowledge in literature by presenting a
relativistic framework of such calculation. 

This paper is organized as follows. 
In section II, we show the integrals of the density and 
the vorticity in the general relativity, in the simple manner 
using the Lagrangian condition. 
This fact suggests strongly that the Lagrangian condition allows us 
to formulate the Lagrangian description in the relativistic cosmology. 
Under this condition, section III presents a perturbative 
Lagrangian approach.  
Summary and Discussion are given in section VI. 
For comparison, the vorticity in the Newtonian cosmology is discussed 
in the appendix A. 
Residual gauge freedoms in the Lagrangian condition are clarified 
in the appendix B. 
Greek indices run from $0$ to $3$, and Latin indices from $1$ to $3$. 
We use the unit, $c = 1$.

\section{Lagrangian description of dust fluid}
 
Let us consider a simple dust universe, in which the energy momentum tensor 
is written as 
\begin{equation}
T^{\mu\nu}=\rho u^{\mu} u^{\nu} . 
\end{equation}
The conservation law $T^{\mu\nu}_{\ \ ;\nu}=0$ gives
\begin{eqnarray}
\rho_{;\mu} u^{\mu} + \rho u^{\mu}_{\ ;\mu} &=& 0 ,   \label{continuity} \\
 u^{\mu}{}_{;\nu}u^{\nu} &=& 0 , \label{geodesics} 
\end{eqnarray}
which are called as the continuity equation and the geodesic equation, 
respectively. 

The vorticity $\omega^{\mu}$ of the fluid flow is defined by \cite{Ehlers} 
\begin{equation}
\omega^{\alpha}\equiv {1 \over 2} \epsilon^{\alpha\mu\nu\rho}
  u_{\mu}u_{\nu;\rho} , 
\label{vorticity}
\end{equation} 
where $\epsilon^{\alpha\mu\nu\rho}$ denotes the complete
anti-symmetric tensor with 
$\epsilon^{0123} = 1/\sqrt{- g}$ and $g\equiv\det(g_{\mu\nu})$.  
 From the geodesic equation Eq. (\ref{geodesics}), we obtain the
propagation equation for the vorticity \cite{Ehlers}:  
\begin{equation}
\omega^{\mu}{}_{;\nu} u^{\nu}+u^{\nu}{}_{;\nu} \omega^{\mu} 
= u^{\mu}{}_{;\nu} \omega^{\nu} . 
\label{eqomegaprop} 
\end{equation}
Using Eq. ($\ref{continuity}$), we have
\begin{eqnarray}
\Bigl( {\omega^{\mu} \over \rho} \Bigr){}_{;\nu} u^{\nu} 
=u^{\mu}{}_{;\nu} \Bigl( {\omega^{\nu} \over \rho} \Bigr) , 
\label{beltrami}
\end{eqnarray}
which may be called as the relativistic Beltrami equation
(cf. Eq. (\ref{eqBeltrami}) in Appendix A). 

The Einstein equations are decomposed with respect to the fluid flow
as follows: 
\begin{eqnarray}
G_{\mu\nu}u^{\mu}u^{\nu}&=&8\pi G \rho , 
\label{H} \\
G_{\mu\nu}u^{\mu}P^{\nu}_{\ \alpha}&=&0 , 
\label{M} \\
G_{\mu\nu}P^{\mu}_{\ \alpha}P^{\nu}_{\ \beta}&=&0 , 
\label{E} 
\end{eqnarray}
where  $P^{\mu}_{\ \nu}$ is the projection tensor 
\begin{equation}
P^{\mu}_{\ \nu}\equiv \delta^{\mu}_{\ \nu}+u^{\mu}u_{\nu} . 
\end{equation}

So far the treatment is fully covariant. In the following, we adopt the 
Lagrangian condition (e.g. \cite{Friedrich}), in which the components of 
the matter 4-velocity take the values of 
\begin{equation}
u^{\mu}=(1,0,0,0) . 
\label{lagrangecon}
\end{equation}
Under this condition, we immediately have $g_{00}=-1$ 
and $u_{\mu} = (-1, g_{0i})$. 
Furthermore, since the geodesic equation Eq. (\ref{geodesics}) 
under the Lagrangian condition simply tells $u_{\mu, 0} = 0$, 
$u_i (= g_{0i})$ are functions of spatial coordinates only:
\begin{equation}
u_i = u_i(\bbox{x}), \quad \mbox{and}\quad g_{0i} = g_{0i}(\bbox{x}). 
\end{equation}
In the Lagrangian description, the continuity equation
(\ref{continuity}) is simply 
\begin{equation}
(\rho \sqrt{-g})_{,0} = 0 . \label{continuity2}
\end{equation}
Therefore, 
\begin{equation}
\rho(\bbox{x},t) = 
 \sqrt{ \frac{g(\bbox{x},t_0)}{g(\bbox{x},t)} } \rho(\bbox{x},t_0) .
\label{masscon}
\end{equation}
Under the Lagrangian condition, the determinant of the metric tensor
can be expressed as
\begin{equation}
g = -(1+\gamma^{ij}g_{0i}g_{0j}) \det (g_{ij}) ,  
\end{equation}
where $\gamma^{ij}$ is the inverse of the spatial metric $g_{ij}$. 
The relativistic Beltrami equation (\ref{beltrami}) also becomes
simply
\begin{equation}
\Bigl( {\omega^{\mu} \over \rho} \Bigr){}_{,0}=0 , 
\end{equation}
which is integrated to give 
\begin{equation}
\frac{\omega^{i}}{\rho} =  
\left.\frac{\omega^{i}}{\rho}\right|_{t_0} . 
\label{cauchy} 
\end{equation}
This is also expressed as 
\begin{equation}
\omega^i(\bbox{x},t) = 
 \sqrt{ \frac{g(\bbox{x},t_0)}{g(\bbox{x},t)} } 
 \omega^i(\bbox{x},t_0) .
\label{cauchy2}
\end{equation}
The $\omega^0$ component is not independent of $\omega^i$. Using the
relation 
$ \omega^{\mu} u_{\mu}=0 $,  
 we obtain 
\begin{equation}
\omega^0=g_{0i} \omega^i , 
\end{equation}

The result Eq. (\ref{cauchy}) tells us that the vorticity is coupled
to the density enhancement and vice versa. In particular, if the vorticity 
does not vanish exactly at an initial time, the vorticity will blow up 
as the density grows larger and larger (i.e. in the collapsing region), 
even if it has only the decaying mode in the linear perturbation theory. 
It should also be emphasized that our results Eqs. (\ref{masscon}) and 
(\ref{cauchy}) in the fully general relativistic treatment precisely
correspond to those in the Newtonian case (see Eqs. (\ref{Nmasscon}) 
and (\ref{Ncauchy}) in the Appendix).

\section{Perturbative Lagrangian approach} 

In the previous section, we solved exactly the equations for the
density and the vorticity. The results Eqs. (\ref{masscon}) and
(\ref{cauchy2}) show that $\rho$ and $\omega^i$ 
are completely written in terms of the determinant of the metric
tensor and their initial values. 
In this section, we solve the metric perturbatively. 
We assume that the background is spatially flat
Friedmann-Lema\^itre-Robertson-Walker (FLRW) universe. 
The extension to the spatially non-flat case must be a straightforward task. 
The perturbed metric is decomposed into 
\begin{eqnarray}\label{permet}
g_{0i}&=& B_{,i}(\bbox{x}) + b_{i}(\bbox{x}) , \\
g_{ij}&=& a^2 \left(\delta_{ij} + 
         2 H_L \delta_{ij} + 2H_{T}{}_{,ij}+
          (h_{i,j}+h_{j,i}) + 2 H_{ij} \right) , \nonumber
\end{eqnarray}
where $B$, $H_L$, and $H_T$ are scalar mode quantities, 
$b_i$ and $h_i$ are the vector (transverse) mode, and $H_{ij}$
is the tensor (transverse-traceless) mode satisfying 
\begin{eqnarray}
b^i_{\ ,i}&=&0 , \\
h^i_{\ ,i}&=&0 , \\
H^i_{\ i}&=&0 , \\
H^{ij}_{\ \ ,j} &=&0 .  
\end{eqnarray}
Raising and lowering indices of the perturbed quantities are done 
by $\delta^{ij}$ and $\delta_{ij}$. 
Using the perturbed metric Eq. (\ref{permet}), the general expression for
the perturbed Einstein tensor up to the linear order is as follows:
\begin{eqnarray}
G_{\mu\nu} u^{\mu} u^{\nu} &=&
3\Bigr(\frac{\dot{a}}{a}\Bigr)^2 
 + 2 \frac{\dot{a}}{a}(3\dot{H}_L + \nabla^2 \dot{H}_T)
- \frac{2}{a^2}\Bigl(\frac{\dot{a}}{a}\nabla^2 B + \nabla^2 H_L\Bigr) ,
\label{eqG00}
\end{eqnarray}
\begin{eqnarray}
G_{\mu\nu} u^{\mu} P^{\nu}_{\ i} &=&
 -2 \Bigl(H_{L} + \frac{\dot{a}}{a}B \Bigr)^{\bbox{\cdot}}{}_{,i} 
 + \frac{1}{2}\nabla^2 \Bigl(\dot{h}_i - \frac{1}{a^2} b_i
       \Bigr) - 2 \Bigl(\frac{\dot{a}}{a}\Bigr)^{\bbox{\cdot}} b_i ,
\label{eqG0i}
\end{eqnarray}
\begin{eqnarray}\label{eqGij}
G_{\mu\nu} P^{\mu}_{\ i} P^{\nu}_{\ j} &=& 
-a^2 \left[ \nabla^2 \Bigl(\ddot{H}_T   
  + 3\frac{\dot{a}}{a} \dot{H}_T
  - \frac{1}{a^2}  H_L 
  - \frac{\dot{a}}{a^3} B \Bigr) 
+ 2 \Bigl( \ddot{H}_L + 3 \frac{\dot{a}}{a}\dot{H}_L \Bigr)  
\right]  \delta_{ij} \\
&& + a^2 \Bigl( \ddot{H}_T + 3 \frac{\dot{a}}{a}\dot{H}_T 
   - \frac{1}{a^2}H_L - \frac{\dot{a}}{a^3}B\Bigr){}_{,ij} \nonumber\\
&&+ \frac{a^2}{2} \left[ 
\Bigl(\ddot{h}_i + 3
\frac{\dot{a}}{a}\dot{h}_i-\frac{\dot{a}}{a^3}b_i\Bigr){}_{,j}
+ \Bigl(\ddot{h}_j + 3
\frac{\dot{a}}{a}\dot{h}_j-\frac{\dot{a}}{a^3}b_j\Bigr){}_{,i}\right] 
                                                 \nonumber\\
&&+ a^2 \Bigl( 
\ddot{H}_{ij} + 3\frac{\dot{a}}{a}\dot{H}_{ij}
-\frac{1}{a^2}\nabla^2 H_{ij} \Bigr), \nonumber
\end{eqnarray}
where an overdot ($\dot{}$) denotes $\partial/\partial t$,
$\nabla^2=\delta^{ij}\partial_i \partial_j$, and we have used
the fact $a(t) = t^{2/3}$ so that 
$G_{\mu\nu} P^{\mu}_{\ i} P^{\nu}_{\ j}$ does not have the background 
quantity.

\subsection{Scalar perturbations}

Using Eqs. (\ref{eqG0i}) and (\ref{eqGij}), the
Einstein equations for the scalar perturbations are
\begin{eqnarray}
\Bigl(H_L + \frac{\dot{a}}{a} B \Bigr)^{\bbox{\cdot}} &=& 0 , 
\label{eqscalar1} \\
\ddot{H}_L + 3 \frac{\dot{a}}{a} \dot{H}_L &=& 0, 
\label{eqscalar2}  \\
\ddot{H}_T + 3 \frac{\dot{a}}{a} \dot{H}_T &=& \frac{1}{a^2} 
\Bigl( H_L + \frac{\dot{a}}{a} B\Bigr) . 
\label{eqscalar3}
\end{eqnarray}
In order to solve the above equations, we use the residual gauge
freedom (see Eq. (\ref{gaugeB}) in Appendix B) to set $B = 0$. 
Then, from Eq. (\ref{eqscalar1}) we have 
\begin{equation}
H_L = H_L(\bbox{x}) \equiv \frac{10}{9} \Psi(\bbox{x}) . 
\end{equation}
Next, from Eq. (\ref{eqscalar3}), we obtain two independent
solutions for $H_T$: one is proportional to $t^{2/3}$ and the other to
$t^{-1}$. According to Eq. (\ref{gaugeHT}), we may also use the
residual gauge freedom for $H_T$ to add any function which does not
depend on $t$. One may choose it as 
\begin{equation}
H_T= \left( t^{2/3}-t_0^{2/3}\right) \Psi(\bbox{x})
      + \left(t^{-1} - t_0^{-1}\right) \Phi(\bbox{x}) ,   
\end{equation}
so that $H_T$ vanishes at the initial time $t=t_0$. 
The initial density field $\rho(\bbox{x}, t_0)$ is also expressed by
the metric. If the density contrast $\delta \equiv
(\rho-\rho_b)/\rho_b$ is sufficiently small, we obtain from
Eqs. (\ref{H}) and (\ref{eqG00})
\begin{equation}
\delta = - \nabla^2 \left( t^{2/3} \Psi + t^{-1} \Phi \right) .
\end{equation}
Therefore, the initial density can be related with the initial 
metric perturbation as 
\begin{equation}
\rho(\bbox{x}, t_0) = \rho_b(t_0) \left[ 
1  - \nabla^2 \left(t_0^{2/3} \Psi + t_0^{-1} \Phi
              \right) \right] .
\end{equation}
The above expression is used only to relate the initial density
fluctuation to the initial metric perturbations in linear regime. Once
the initial seed is given, the later non-linear evolution is evaluated 
by the non-perturbative expression Eq.~(\ref{masscon}).

\subsection{Vector perturbations}
The Einstein equations for the vector perturbations are 
\begin{eqnarray}
\nabla^2 \Bigl( \dot{h}_i - \frac{1}{a^2}b_i\Bigr) &=& 
4 \Bigl(\frac{\dot{a}}{a}\Bigr)^{\bbox{\cdot}} b_i ,
\label{eqvector1}\\
\ddot{h}_i + 3 \frac{\dot{a}}{a} \dot{h}_i 
 - \frac{\dot{a}}{a^3} b_i &=& 0 . 
\label{eqvector2}
\end{eqnarray}
Introducing $\beta_i$ as 
\begin{equation}
b_i(\bbox{x})=\nabla^2 \beta_i(\bbox{x}) , 
\end{equation}
Eq. (\ref{eqvector1}) is solved to give 
\begin{equation}
h_i=-3 \left(t^{-1/3}- t_0^{-1/3}\right) \nabla^2 \beta_i(\bbox{x})
    +{8 \over 3} \left(t^{-1}-t_0^{-1}\right) \beta_i(\bbox{x}) ,
\label{vector}
\end{equation}
where we again used the residual gauge freedom
(cf. Eq. (\ref{gaugehi}) in the Appendix B) to set
$h_i(\bbox{x},t_0) = 0$. 
The $\beta_i(\bbox{x})$ is directly related to the 
initial value of the vorticity field.   From Eq. (\ref{cauchy}),
we have 
\begin{equation}
\omega^i(\bbox{x},t_0) = 
 \frac{1}{2\sqrt{-g(\bbox{x},t_0)}} 
\varepsilon^{ijk} \nabla^2 \beta_{j,k}, 
\end{equation}
where $\varepsilon^{ijk}$ is the 3-dimensional Levi-Civita symbol with
$\varepsilon^{123} = 1$. 
Particularly when the deviation from the background is sufficiently
small, we obtain the first-order expression 
\begin{equation}
\omega^i(\bbox{x},t_0) \simeq  
 \frac{1}{2 a^3(t_0)} 
\varepsilon^{ijk} \nabla^2 \beta_{j,k}.  
\end{equation}

\subsection{Tensor perturbations}

The equation for the tensor perturbations is 
\begin{equation}
\ddot{H}_{ij} + 3\frac{\dot{a}}{a}\dot{H}_{ij}
-\frac{1}{a^2}\nabla^2 H_{ij} = 0 . 
\end{equation}
This is a homogeneous wave equation in the expanding universe. 
The solutions are well known and we will not discuss the detail here. 
See, e.g., \cite{Weinberg}. 

Before closing this section, we emphasize the following point: 
As for the metric, our result is identical to that of the linear 
perturbation theory (e.g. \cite{Bardeen}). However, there is an important
difference.  In Bardeen's paper, it is essential to linearize 
the density contrast. On the other hand, our Lagrangian approach does not 
rely on the assumption that the density contrast should be small. 
It is actually an important advantage of the Lagrangian approach 
that it uses (or extrapolates) the well-known solutions of the linear 
theory to express the non-linear density contrast.

\section{Summary and discussion}

Motivated by the fact that the Lagrangian condition enables us to
obtain simply the integrals of the density and the vorticity along the
fluid flow, we have developed the Lagrangian perturbation theory in
the general relativistic cosmology, fully using this condition. The
main advantage of the present Lagrangian theory is to be amenable to
the vorticity, while previous works are limited within the
irrotational fluid. In this approach, only the metric is expanded and
its behaviors are determined perturbatively through the evolution
equations, while the density and the vorticity are calculated
non-perturbatively from the integrals along the flow. Seeds of the
density and vorticity fluctuations can be related with the metric
perturbation at an initial time, only when the constraint equation is
solved as the Poisson-like equation. Hence, this simple approach is
the first that can describe the dynamics of the dust fluid with the
rotational motion as well as the non-linear density fluctuations, up
to the caustic formation.

As an illustration, the first order Lagrangian perturbation 
has been solved explicitly. 
It is natural that at the first order of the metric perturbation, 
our result agrees with that of the usual linear theory \cite{Bardeen} 
when we assume the small density contrast. 
Contrary to the standard linear perturbation theory, it is explicitly
shown in our approach that the
vorticity is coupled to the density contrast and is amplified in high
density (i.e., collapsing) region. 

In this paper, we have presented a relativistic framework to describe
non-linear evolution of the density fluctuation which is not
restricted to the irrotational case. Once the framework is given, it
should be straightforward to apply it to higher-order
computations. Actually, an extension to the second order is being
investigated by Morita \cite{Morita}. More comprehensive study of
non-linear couplings will be of the subject of future investigation.

\section*{Acknowledgment}

H.A. is indebted to K. Tomita for hospitality at Yukawa Institute 
for the Theoretical Physics, where a part of this work was done.
MK would like to thank Gerhard B\"orner for hospitality at
Max-Planck-Institut f\"ur Astrophysik, where a part of this work was
done. 

\appendix
\section{Vorticity in the Newtonian cosmology} 

In this Appendix, we summarize the treatment of vorticity in the Newtonian 
cosmology. Using expanding coordinates  
\( \bbox{r} = a(t) \bbox{x} \), 
the continuity equation and the equation of motion are
\begin{equation}\label{eqcont}
\frac{\partial \rho}{\partial t} + 3 \frac{\dot{a}}{a} \rho
 + \frac{1}{a}\nabla \cdot (\rho \bbox{v}) = 0 , 
\end{equation}
\begin{equation}\label{eqmot}
\frac{\partial \bbox{v}}{\partial t} + \frac{\dot{a}}{a}\bbox{v}
 + \frac{1}{a} (\bbox{v}\cdot\nabla) \bbox{v} = -\frac{1}{a}\nabla\phi , 
\end{equation}
where \(\rho\) is the density of dust matter, \(\bbox{v}\) is the
peculiar velocity defined by
\begin{equation}
\bbox{v} \equiv \frac{d\bbox{r}}{dt} - \frac{\dot{a}}{a}\bbox{r} 
= a \frac{d\bbox{x}}{dt} ,  
\end{equation}
and \( \phi \) is the gravitational potential which
satisfies the Poisson equation 
\begin{equation}
\nabla^2 \phi =  4 \pi G a^2 (\rho - \rho_b). 
\end{equation}
Let us introduce the Lagrangian time derivative 
\begin{equation}
\frac{d}{dt} \equiv \frac{\partial}{\partial t} + 
 \frac{1}{a}\bbox{v}\cdot\nabla , 
\end{equation}
and the Lagrangian coordinates \( \bbox{q}\equiv \bbox{x}(t_0) \). 
In the Lagrangian description, the continuity equation \
(\ref{eqcont}) is
\begin{equation}
\frac{d\rho}{dt} + 3 \frac{\dot{a}}{a}\rho 
 + \frac{\rho}{a}\nabla \cdot \bbox{v} = 0 , 
\label{Ncontinuity2}
\end{equation}
which is solved to give
\begin{equation}\label{Nmasscon}
\rho = \left(\frac{a_0}{a}\right)^3 \frac{\rho_0}
       {\det \left( \partial x^i / \partial q^j \right)} .
\end{equation}
We now discuss the vortical part of the fluid flow. The vorticity
 is defined by
\begin{equation}
\bbox{\omega} \equiv \frac{1}{a} \nabla \times \bbox{v} . 
\end{equation}
{}From Eq. $(\ref{eqmot})$, we obtain the propagation equation for the
vorticity: 
\begin{equation}
\frac{d\bbox{\omega}}{dt} + 2 \frac{\dot{a}}{a}\bbox{\omega}
 + \frac{\bbox{\omega}}{a}\nabla\cdot\bbox{v} 
 = \frac{1}{a}(\bbox{\omega}\cdot\nabla)\bbox{v}.  
\end{equation}
Using Eq.\ (\ref{Ncontinuity2}), we accordingly have the diffusion
equation of Beltrami: 
\begin{equation}\label{eqBeltrami}
\frac{d}{dt}\left(\frac{\bbox{\omega}}{a \rho}\right) = 
  \left(\frac{\bbox{\omega}}{a \rho}\cdot\nabla \right) \bbox{v}.  
\end{equation}
It is instructive to define the vector components of the vorticity in
the Lagrangian frame as follows: 
\begin{equation}
\tilde{\omega}^i \equiv \frac{\partial q^i}{\partial x^j} \omega^j . 
\end{equation}
Then, Beltrami's equation (\ref{eqBeltrami}) simply reads
\begin{equation}
\frac{d}{dt}\left(\frac{\tilde{\bbox{\omega}}}{a \rho}\right) =
\bbox{0},  
\end{equation}
which is manifestly solved to give 
\begin{equation}\label{Ncauchy}
\frac{\tilde{\bbox{\omega}}}{a \rho} =  
\left.\frac{\bbox{\omega}}{a \rho}\right|_{t_0} , 
\end{equation}
or
\begin{equation}\label{eqCauchy}
\omega^i(\bbox{q},t) = \frac{a \rho}{a_0 \rho_0} 
\frac{\partial x^i}{\partial q^j} \omega^j(\bbox{q},t_0) . 
\end{equation}
The result Eq. (\ref{eqCauchy}) is known as Cauchy's integral \cite{Buchert}. 


\section{Residual gauge freedom in the Lagrangian condition}

The general gauge transformation to the first order is induced 
by the infinitesimal coordinate transformation 
\begin{equation}
\tilde{x}^{\mu}=x^{\mu}+\xi^{\mu} . 
\end{equation}
The changes due to the gauge transformation are
\begin{eqnarray}
\delta_{\xi} g_{\mu\nu} 
&=& -g_{\mu\nu,\alpha}\xi^{\alpha}-g_{\mu\alpha}\xi^{\alpha}{}_{,\nu}
-g_{\nu\alpha}\xi^{\alpha}{}_{,\mu} , 
\label{deltag} \\
\delta_{\xi}u^{\mu} 
&=&\xi^{\mu}{}_{,\nu}u^{\nu}-u^{\mu}{}_{,\nu}\xi^{\nu} . 
\label{deltau}
\end{eqnarray}

For the spatially flat background, $\xi^{\mu}$ are decomposed into each 
mode, 
\begin{equation}
\xi^{\mu}=(T,\delta^{ij} L_{,j}+\ell^i) , 
\end{equation}
where the vector mode quantity $\ell^i$ satisfies $\ell^i_{\, ,i} = 0$. In
order to maintain the Lagrangian condition $u^{\mu} = (1,0,0,0)$, we
have $\delta_{\xi}u^{\mu} = 0$, which leads to $\xi^{\mu}{}_{,0}= 0$. 
Therefore, 
\begin{equation}
T = T(\bbox{x}),\quad  L_{,i} =L_{,i}(\bbox{x}), \quad
\ell_i = \ell_i(\bbox{x}).  
\label{gaugeTL} 
\end{equation}
The changes of the metric tensor due to the residual gauge freedom are
\begin{eqnarray}
\delta_{\xi} g_{0i} &=& T_{,i} , \\
\delta_{\xi} g_{ij} &=& -a^2 \left(
2 \frac{\dot{a}}{a}T\delta_{ij} + 2 L_{,ij} + \ell_{i,j} + \ell_{j,i}
\right). 
\end{eqnarray} 
Hence we obtain 
\begin{eqnarray}
\tilde{B} &=& B + T(\bbox{x}), \label{gaugeB}\\
\tilde{H}_L &=& H_L -  \frac{\dot{a}}{a} T(\bbox{x}), \label{gaugeHL}\\
\tilde{H}_T &=& H_T - L(\bbox{x}), \label{gaugeHT}\\
\tilde{h}_i &=& h_i - \ell_i(\bbox{x}). \label{gaugehi} 
\end{eqnarray}
The vector mode quantity $b_i$ and the tensor mode quantity
$H_{ij}$ are gauge-invariant.

\end{document}